\documentstyle[12pt,fleqn]{article}
\def\One{\mbox{{\sf 1}\kern-0.25em{\bf l}}}
\def\beq{\begin{equation}}
\def\eeq{\end{equation}}

\begin{document}
\vspace*{15mm}
\begin{center}
{\Large Higher order Schmidt decompositions} \\[1cm]
Asher Peres\\[7mm]
{\sl Department of Physics, Technion -- Israel Institute of
Technology, 32 000 Haifa, Israel}\\[1cm]
\end{center}\vspace{2cm}

\noindent{\bf Abstract}\bigskip

Necessary and sufficient conditions are given for the existence of
extended Schmidt decompositions, with more than two subspaces.\\[2cm]

Schmidt's theorem [1, 2] has a fundamental importance in the quantum
theory of measurement. In the simple case of finite dimensional spaces
[3, 4], the theorem asserts that any double sum
\beq \Psi=\sum_{mn} A_{mn}\,x_m\,y_n\,,\eeq
can be converted into a single sum
\beq \Psi=\sum_\mu a_\mu\,\xi_\mu\,\eta_\mu\,, \eeq
by means of {\em unitary\/} transformations
\beq \xi_\mu=\sum_m U_{\mu m}\,x_m\qquad\qquad{\rm and}
  \qquad\qquad \eta_\mu=\sum_n V_{\mu n}\,y_n\,. \eeq
If $\{x_m\}$ and $\{y_n\}$ are two orthonormal bases for two distinct
vector spaces, then $\{\xi_\mu\}$ and $\{\eta_\mu\}$ also are two,
possibly incomplete, orthonormal vector bases for these two spaces. The
absolute values $|a_\mu|$ are called the {\em singular values\/} [3] of
the matrix $A$, and are easily calculated by noting that $|a_\mu|^2$ are
the nonvanishing eigenvalues of the Hermitian matrices $AA^\dagger$ and
$A^\dagger A$. The corresponding sets of eigenvectors are $\{\xi_\mu\}$
and $\{\eta_\mu\}$, respectively.

A natural question is whether this process can be extended to more than
two subspaces. Such an extension of Schmidt's theorem would be useful
for modal interpretations of quantum theory, and triple sums can indeed
sometimes be found in the literature on that subject [5--7]. Although
none of these last references is technically wrong, because of the
context where these triple sums appear, they may give the impression
that multiple Schmidt decompositions are always possible. I have
actually seen such a false assumption used in a preprint, already
accepted for publication, whose author will be eternally grateful to me
for warning him of the error before it was too late.

The unlikeliness of occurence of multiple Schmidt decompositions can
readily be seen by counting the free parameters involved: for example,
if there are three particles, each of which described by a
$d$-dimensional space, their combined (pure) state, in a $d^3$-dimensional
space, depends on $2(d^3-1)$ real parameters (after discarding overall
normalization and phase factors). On the other hand, the three
unimodular unitary transformations which can be performed for these
three particles have only $3(d^2-1)$ free parameters, not enough to
solve the problem in general.

This negative result prompts a more difficult question: what are the
necessary and sufficient conditions for the occurence of a triple, or
multiple, Schmidt decomposition? Consider a triple sum
\beq \Psi=\sum_{mns} A_{mns}\,x_m\,y_n\,z_s\,,\eeq
where $\{x_m\}$, $\{y_n\}$ and $\{z_s\}$ are three orthonormal bases in
three distinct vector spaces. Can we rewrite the above expression as
$\sum a_\mu\xi_\mu\eta_\mu\zeta_\mu$, with three orthonormal sets,
$\{\xi_\mu\}$, $\{\eta_\mu\}$ and $\{\zeta_\mu\}$? Schmidt's theorem
only asserts that Eq.~(\theequation) can be unitarily converted into
\beq \Psi=\sum_\mu a_\mu\,\xi_\mu\,\omega_\mu\,, \label{a}\eeq
where $\{\xi_\mu\}$ is a set of orthonormal vectors in the space that
was spanned by $\{x_m\}$, and $\{\omega_\mu\}$ is a set of orthonormal
vectors in the space spanned by the set $\{y_n\otimes z_s\}$.

Consider first, for simplicity, the case where all the $|a_\mu|$ are
different, so that the decomposition (\theequation) is unique. We can
then write, if a triple Schmidt decomposition exists,
\beq \forall\mu\qquad\omega_\mu=
  \sum_{ns}{\mit\Omega}_{\mu ns}\,y_n\,z_s=\eta_\mu\,\zeta_\mu\,.\eeq
This implies that all the ${\mit\Omega}_\mu$ matrices, namely the
matrices with elements $({\mit\Omega}_\mu)_{ns}={\mit\Omega}_{\mu ns}$,
are of rank~1, and satisfy both
${\mit\Omega}^\dagger_\mu\,{\mit\Omega}_\nu=0$ and
${\mit\Omega}_\mu\,{\mit\Omega}^\dagger_\nu=0$ if $\mu\neq\nu$. These
conditions are obviously necessary. They are also sufficient, because if
they hold, we can obtain orthogonal sets $\{\eta_\mu\}$ and
$\{\zeta_\mu\}$ as the eigenvectors of
${\mit\Omega}^\dagger_\mu\,{\mit\Omega}_\mu$ and
${\mit\Omega}_\mu\,{\mit\Omega}^\dagger_\mu$, respectively.

If several nonvanishing $|a_\mu|$ are equal, the decomposition (\ref{a})
is not unique, and the situation becomes more complicated. Suppose that
$n$ of these $|a_\mu|$ are equal, so that the corresponding $\xi_\mu$
and $\omega_\mu$ are defined only up to an arbitrary unitary
transformation, represented by a matrix of order $n$. Consider the subspace
spanned by these $\omega_\mu$.
Then, obviously, the corresponding ${\mit\Omega}_\mu$ matrices should be
of a rank not exceeding $n$. Moreover, there should be $n$ linear
combinations of the ${\mit\Omega}_\mu$ matrices, generated by the
inverse of the above unitary transformation,
such that the resulting matrices, ${\mit\Omega}'_\mu$ say, are all of rank~1,
and satisfy
orthogonality conditions as specified above for the nondegenerate case,
namely ${\mit\Omega}'^\dagger_\mu\,{\mit\Omega}'_\nu=0$ and
${\mit\Omega}'_\mu\,{\mit\Omega}'^\dagger_\nu=0$ if $\mu\neq\nu$. This
completes the solution of this problem.

\bigskip I am grateful to Leslie Ballentine for drawing my attention to
this important issue [8] and encouraging me to publish these results.
This work was supported by the Gerard Swope Fund, and the Fund for
Encouragement of Research.\bigskip

\frenchspacing
\begin{enumerate}
\item E. Schmidt, Math. Ann. 63 (1907) 433.
\item F. Riesz and B. Sz.-Nagy, Functional analysis (Ungar, New York,
1955) p. 243.
\item P. Lancaster and M. Tismenetsky, The theory of matrices (Academic
Press, Boston, 1985) p. 182.
\item A. Peres, Quantum theory: concepts and methods (Kluwer, Dordrecht,
1993) p. 123.
\item S. Kochen, in ``Symposium on the foundations of modern physics''
ed.~by P.~Lahti and P.~Mittelstaedt (World Scientific, Singapore, 1985)
p.~151.
\item E. J. Squires, in ``Quantum theory without reduction'' ed. by
M.~Cini and \mbox{J.-M.} L\'evy-Leblond (Hilger, Bristol, 1990) p.~151.
\item R. Healey, Found. Phys. Letters 6 (1993) 37.
\item L. E. Ballentine, Am. J. Phys. 63 (1995) 285.
\end{enumerate}

\end{document}